\documentclass[useAMS,usenatbib]{mn2e}
\usepackage{graphicx}
\usepackage{epstopdf}
\usepackage{amsmath}
\usepackage{amssymb}
\title[Wind from hot accretion flows]{On the wind production from hot accretion flows with different accretion rates}
\author[Bu \& Gan]{De-Fu Bu\thanks{E-mail: dfbu@shao.ac.cn (DB)} and Zhao-Ming Gan \\
Key Laboratory for Research in Galaxies and Cosmology, Shanghai Astronomical Observatory, Chinese Academy of Sciences,\\ 80 Nandan Road, Shanghai 200030, China}
\begin{document}


\pagerange{\pageref{firstpage}--\pageref{lastpage}} \pubyear{2002}

\maketitle

\label{firstpage}
\begin{abstract}
 We perform two-dimensional simulations to study how the wind strength changes with accretion rate. We take into account bremsstrahlung, synchrotron radiation and the Comptonization. We find that when the accretion rate is low, radiative cooling is not important, the accretion flow is hot accretion flow. For the hot accretion flow, wind is very strong. The mass flux of wind can be $\sim 50\%$ of the mass inflow rate. When the accretion rate increases to a value at which radiative cooling rate is roughly equal to or slightly larger than viscous heating rate, cold clumps can form around the equatorial plane. In this case, the gas pressure gradient force is small and wind is very weak. Our results may be useful for the sub-grid model of active galactic nuclear feedback study.
\end{abstract}

\begin{keywords}
accretion, accretion discs -- black hole physics -- hydrodynamics.
\end{keywords}
\section{Introduction}
In the 1990s, people used analytical method to investigate the dynamics and spectral of hot accretion flows (Narayan \& Yi 1994; Narayan \& Yi 1995, hereafter NY95; Abramowicz et al. 1995; Kato et al. 1998; Narayan et al. 1998). The black holes in
low-luminosity active galactic nuclei (LLAGNs; e.g. Ho 2008; Antonucci 2012; Done 2014) accrete gas in hot accretion flow mode. Hot accretion flow is also operating in the
hard/quiescent states of black hole X-ray binaries (e.g. Esin et al. 1997; Fender et al. 2004; Zdziarski \& Gierli\'{n}ski 2004; Narayan 2005; Remillard \& McClintock 2006; Narayan \& McClintock 2008; Belloni 2010; Wu et al. 2013; Yuan \&
Narayan 2014). Since the end of the 20th century, many simulations have been carried out to study the dynamics of hot accretion flow (e.g. Stone et al. 1999; Igumenshchev \& Abramowicz 1999, 2000; Hawley et al. 2001; Machida et al. 2001; Stone \& Pringle 2001; Hawley \& Balbus 2002; De Villiers et al. 2003; Igumenshchev et al. 2003; Pen et al. 2003; Beckwith et al. 2008; Pang et al. 2011; Tchekhovskoy et al. 2011; Tchekhovskoy \& McKinney 2012; Yuan et al. 2012a, 2012b; McKinney et al. 2012; Narayan et al. 2012; Sadowski et al. 2013; Mo\'{s}cibrodzka et al. 2014).

An very important result found by numerical simulations is that strong wind exists in hot accretion flow (e.g. Yuan et al. 2012b; Yuan et al. 2015; Narayan et al. 2012; Li et
al. 2013). In the hydrodynamic (HD) simulation, there is no magnetic field, wind is found to be driven by the combination of gas pressure gradient and centrifugal forces. In the magneto-hydrodynamic (MHD) simulation, it is found that wind is driven by the combination of gas pressure gradient, magnetic pressure gradient and centrifugal forces (Yuan et al. 2015, see also Moller \& Sadowski
2015). Recently, observations of both LLAGNs (e.g. Crenshaw \& Kramemer 2012; Tombesi et al. 2010; 2014; Wang et al. 2013; Cheung et al. 2016) and the hard state of black hole X-ray binaries (Homan et al. 2016) show that winds are present in hot accretion flow. The reason for the existence of wind is also investigated by analytical works (e.g. Begelman 2012; Gu 2015; Mosallanezhad et al. 2016).
Bu et al. (2016a; 2016b) study whether wind can be produced beyond the Bondi radius. In these studies, the gravity of nuclear star cluster is taken into account. It is found that the radial profile of the gravity becomes flatter when stars gravity is taken into account. Also, it is found that wind can not be produced in the presence of star gravity. Wind production can only occur within the
Bondi radius of hot accretion flow.

The simulations introduced above neglect radiative cooling. Non-radiative simulations of hot accretion flow can only be applied when the black hole accretes at very low rate, $\dot {M}\leq 10^{-6} \dot {M}_{Edd}$, $\dot {M}_{Edd}$ is the Eddington accretion rate (Sadowski et al. 2017). With the increase of mass accretion rate, radiative cooling effects will become important. Strong radiative cooling can decrease the gas pressure. As mentioned above, gas pressure gradient is an important force for driving wind. We can expect that when radiation becomes strong, wind should become weak. In this paper, we study how the wind strength changes with mass accretion rate.

We are interested in wind due to the following two reasons. First, wind can take away mass which can change the density profile of accretion flow. The change of density profile can affect the emitted spectrum of an accretion flow (e.g. Quataert \& Narayan 1999; Yuan et al. 2003). Second, it is now widely believed that AGNs feedback plays an important role in the evolution of their host galaxies (e.g. Ciotti \& Ostriker 1997; 2001; 2007; Proga et al. 2000; Proga 2007; Kurosawa \& Proga 2009; Novak et al. 2011; Gan et al. 2014, 2017; King \& Pounds 2015; Ciotti et al. 2017). Wind can push away the gas outside an AGN. This can reduce the feeding rate of the central black hole. Also, the star formation rate may also be changed (e.g. Ciotti et al. 2017). In order to study wind feedback, we first need to study the strength of wind.

The effects of radiative cooling on hot accretion flow have been taken into account recently (e.g. Ryan et al. 2017; Wu et al. 2016). In these papers, the effects of radiative cooling on wind strength have not been discussed.

Yuan \& Bu (2010) studied how the wind strength changes with mass accretion rate by performing HD simulations. In that work, only bremsstrahlung radiation is taken into account. It is found that when the accretion rate is high, the radiative cooling rate can exceed the viscous heating rate in the outer region (beyond 30 Schwarzschild radius). In this case, the accretion flow outside 30 Schwarzschild radius should be luminous hot accretion flow (LHAF; Yuan 2001). They found that strong wind still exists for LHAF. Note that synchrotron emission and its Comptonization are neglected in Yuan \& Bu (2010).  Synchrotron emission is very important in the inner region of hot accretion flow. If synchrotron emission is included, in the inner region of hot accretion flow, radiative cooling rate may also exceed viscous heating rate. In this case, the strength of winds may change. In this paper, we study how wind strength changes with mass accretion rate by taking into account bremsstrahlung, synchrotron and the Comptonization.


In  \S 2, we will describe the basic equations and the simulation method. In \S 3, we will present the results. We discuss
and summarize our results in \S4.

\section{Numerical method and models }
\subsection{Basic equations}
In hot accretion flows, the temperatures of electrons and ions can be very different (Yuan \& Narayan 2014). The reasons are as follows. First, the electrons can radiate energy much quicker than ions; Second, the ions can cool themselves by transferring their energy to electrons via Coulomb interaction. In hot accretion flow, gas density is low. Thus, Coulomb collision is very inefficient. The ions can not cool efficiently. Finally, more viscous heating energy goes into ions (see the introduction below).  In this paper, we solve two energy equations, i.e., one for ions and one for electrons.

We employ spherical coordinates ($r,\theta,\phi$)
and assume axisymmetric ($\partial/\partial \phi=0$) . We use the
Zeus-3D code (Clarke 1996) to solve the HD
equations below:

\begin{equation}
 \frac{d\rho}{dt} + \rho \nabla \cdot {\bf v} = 0,
\end{equation}
\begin{equation}
 \rho \frac{d{\bf v}}{dt} = -\nabla P_{tot} - \rho \nabla \Phi + \nabla \cdot \bf{T}
\end{equation}
\begin{equation}
 \rho \frac{d(e_i/\rho)}{dt} = -P_{i}\nabla \cdot {\bf v} +(1-\delta) \bf {T}^2/\mu - q^{ie}
\end{equation}
\begin{equation}
 \rho \frac{d(e_e/\rho)}{dt} = -P_{e}\nabla \cdot {\bf v} +\delta \bf {T}^2/\mu + q^{ie} - Q^{-}
\end{equation}
Here, $\rho$ is the mass density, $\bf v$ is the velocity, $P_{tot} \equiv P_{gas}+P_{mag}$ is
the total (gas $+$ magnetic) pressure, ${\bf T}$ is the anomalous stress tensor, $\Phi$ is the gravitational potential.
$e_i(=P_i/(\gamma_i-1))$ and $e_e(=P_e/(\gamma_e-1))$ are the internal energies of ions and electrons. $\gamma_i$ and $\gamma_e$ are the respective adiabatic indices, $P_i$ and $P_e$ are the respective pressures. The ions are always non-relativistic in hot accretion flow, we assume $\gamma_i=5/3$. But for electrons, in the inner regions of the accretion flow, the electrons become relativistic, and we assume $\gamma_e=4/3$ (Yuan \& Narayan 2014). $q^{ie}$ is the rate of transfer of thermal energy from ions to electrons via Coulomb collisions. The parameter $\delta$ denotes the fraction of the viscously dissipated energy that directly heats electrons; the remainder ($1-\delta$) goes into the ions. $Q^-$ is the radiative cooling term.

Recently, Ryan et al. (2017) perform two-temperature simulation to study the radiative efficiency of hot accretion flow with $10^{-5} \dot {M}_{Edd}$. They find that, at such low accretion rate, the radiative efficiency can be $1\%$. Sadowski et al. (2017) also perform two-temperature simulation to study hot accretion flow. They find that when $\dot {M} \sim 4\times 10^{-4} \dot {M}_{Edd}$, the flow behaves very differently from those without considering radiative cooling.

The viscous stress tensor $T$ in Equations (2)-(4) is used to mimic the angular momentum transfer by MHD turbulence associated with tangled magnetic field.
The detailed forms of viscous stress can be found in Stone et al. (1999; Equations (4) and (5)). We adopt the coefficient of shear viscosity $\mu=\nu \rho$ and assume $\nu = \alpha c_s^2/\Omega_k$ ($c_s$ and $\Omega_k$ are sound speed and angular Keplerian velocity), which is the usual $\alpha$ viscosity description. If we use $B_r$ and $B_\phi$ to represent the $r$ and $\phi$ components of magnetic field, the Maxwell stress is $B_rB_\phi/4\pi \sim P_{mag}$. The viscous coefficient $\alpha$ can be expressed by $\alpha = B_rB_\phi/4\pi/P_{gas} \sim P_{mag}/P_{gas}= 1/\beta $ (Blackman et al. 2008; Guan et al. 2009; Sorathia et al. 2012). In all of our models, we assume that $\alpha=1/\beta$.

We use the pseudo-Newtonian potential, $\Phi=-GM/(r-r_s)$, where $G$ is
the gravitational constant, $M$ is the mass of the
central black hole and $r_s$ is the Schwarzschild radius. We set $GM=r_s=1$. We use $r_s$ to
normalize the length-scale. Time is in unit of the Keplerian orbital
time at $200r_s$.

\subsection{Radiative cooling}
The ions and electrons exchange energy through Coulomb collision ($q^{ie}$ in Equations (3) and (4)). The exact expression for $q^{ie}$ can be found in NY95 (see Equation (3.1) in NY95).

In hot accretion flow, the radiation mechanisms are the bremsstrahlung, synchrotron, and the Comptonization. The radiative cooling rate in Equation (4) is:
\begin{equation}
Q^-=Q^-_{brem}+Q^-_{syn}+Q^-_{brem,C}+Q^-_{syn,C}
\end{equation}
Here, $Q^-_{brem}$ is the bremsstrahlung cooling including both the electron-ion and electron-electron collisions (Equation (3.4) in NY95). $Q^-_{syn}$ is the synchrotron emission (Equation (3.17) in NY95).

We do not take into account global Comptonization effects (interaction between photons from inner region and gas in outer region). Instead, we use the Compton enhancement factor $\eta$ (Dermer et al. 1991), which is defined to be the average change in energy of a photon between injection and escape. The radiation cooling rate due to the Comptonized synchrotron and bremsstrahlung emission is,
\begin{equation}
Q^-_{brem,C}+Q^-_{syn,C}=(\eta-1)(Q^-_{brem}+Q^-_{syn})
\end{equation}
The detailed formula of $\eta$ can be found in NY95 (see Equation (3.19) in NY95). Note that the optical depth needed for calculating $\eta$ at any radius is measured from $\theta=0^\circ$ to $180^\circ$.
Note that Comptonization used in this paper is local. Thus, it is not accurate. We will discuss this point in the last section.

\subsection{Initial conditions and numerical settings}
Torus is widely used in simulations as initial condition. There are several torus models. For example, there are torus models in general relativistic potential (e.g., Fishbone \& Moncrief 1976; Abramowicz et al. 1978). There are also torus models in Newtonian potential (e.g. Papaloizou \& Pringle 1984). In this paper,
following Stone et al. (1999), we use the torus introduced by Papaloizou \& Pringle (1984) as our initial condition. The torus has a constant specific angular momentum. The specific angular momentum of the torus is equal to the Keplerian angular momentum at the torus center. The pressure and density in the torus are related by a polytropic equation of state $P=A \rho^\gamma$. {\bf In this paper, we set $A=0.022$.} The equation describing the torus is given by
(Papaloizou \& Pringle 1984),
\begin{equation}
 \frac{P}{\rho} = \frac{GM}{(n+1)R_{0}} \left[ \frac{R_{0}}{r} - \frac{1}{2}
\left( \frac{R_{0}}{r \sin \theta} \right)^{2} - \frac{1}{2d}
\right] .
\end{equation}
Here, $R_0$ is the radius of the torus center (density maximum), $n
= (\gamma -1 )^{-1}$ is the polytropic index and $d$ is the
distortion of the equilibrium torus. In this torus model, we have three free parameters. The first one is the density at the torus center. The second one is the torus center $R_0$. The third one is the distortion $d$ parameter, which determines the thickness of the torus and also the locations of the inner and outer boundaries of the torus. In this paper, we assume that $R_0=200r_s$ and $d=1.125$. Correspondingly the inner and outer boundaries of the torus are located at $150r_s$ and $300r_s$, respectively. The torus is embedded in a low-density medium. The density of the medium is 4 orders of magnitude smaller than the density at the torus center.

In this paper, we assume the black hole mass is $M=10^8M_{\odot}$, with $M_{\odot}$ is the solar mass. The initial torus center is located at $200r_s$. Table 1 lists our models and the results.

The radial computational domain is from $R_{ \rm in} =1.25$
to $R_{\rm out}$ = 1000 $r_{\rm s}$. In $\theta$ direction, we have
$0^\circ < \theta < 180^\circ $.
The radial grids are logarithmically spaced (grid spacing ${\rm d}r
\propto r$). In angular direction, the grids are uniformly spaced.
The axisymmetric boundary conditions are used at $\theta=0^\circ$
and $180^\circ$. We adopt outflow boundary conditions at both the
inner and outer radial boundaries. The standard resolution is $192
\times 88$.

\subsection {Models}

\begin{table*} \caption{Simulation parameters }
\begin{tabular}{ccccccc}
\hline \hline
Name & $\alpha$ & $\delta^\dag$ & $\rho_0/\rho_{max}$ & resolution & $\dot{M}_{\rm BH}/\dot{M}_{\rm Edd}$  & $s^\ddag$ \\
\hline
A0  &  0.1   & 0.1 & $10^{-4}$ & $192\times 88$ & $4\times 10^{-4}$ &  \\
A0lm&  0.1   & 0.1 & $10^{-5}$ & $192\times 88$ & $4\times 10^{-4}$ &  \\
A0h &  0.1   & 0.1 & $10^{-4}$ & $384\times 176$ & $4\times 10^{-4}$ &  \\
A1  &  0.1   & 0.1 & $10^{-4}$ & $192\times 88$ & $5\times 10^{-5}$ & $0.54$ \\
A2  &  0.1   & 0.1 & $10^{-4}$ & $192\times 88$ & $7\times 10^{-6}$ & $0.57$ \\
B0  &  0.1   & 0.5 & $10^{-4}$ & $192\times 88$ & $3.3 \times 10^{-4}$ &  \\
B1  &  0.1   & 0.5 & $10^{-4}$ & $192\times 88$ & $5.2 \times 10^{-5}$ & $0.56$ \\
B2  &  0.1   & 0.5 & $10^{-4}$ & $192\times 88$ & $4\times 10^{-6}$ & $0.53$ \\
C0  &  0.01  & 0.1 & $10^{-4}$ & $192\times 88$ & $2.2\times 10^{-5}$ &  \\
C1  &  0.01  & 0.1 & $10^{-4}$ & $192\times 88$ & $6\times 10^{-6}$ & $0.55$ \\
C2  &  0.01  & 0.1 & $10^{-4}$ & $192\times 88$ & $5\times 10^{-7}$ & $0.6$ \\
\hline
\end{tabular}

$\delta^\dag$ is the fraction of the viscously dissipated energy that directly heats electrons. \\
$\dot{M}_{\rm BH}$ is the accretion rate measured at the inner boundary of the simulation. \\
$s^\ddag$ is the power law index of the mass inflow rate measured beyond $10r_s$. In modes A0, A0lm, A0h, B0 and C0, a cold disk or cold clumps form at the equatorial plane. There is no quasi-steady state for these models. Therefore, there is no $s$ value for these models.
\end{table*}

Table 1 lists the main parameters and results of our models. Column (2) gives the $\alpha$ value of the viscosity. The viscous heating can heat both ions and electrons. In this paper, we assume that the ratio of the viscous heating energy that goes into electrons to the total viscous heating rate is $\delta$. Thus, the left of $1-\delta$ times the viscous heating energy goes into ions. Column (3) shows the value of $\delta$. Column (4) shows the ratio of the background density where the torus embeds to the density at the torus center. Column (5) shows the resolution. Column (6) is the mass accretion rate at the inner boundary. Column (7) shows the power law index of the mass inflow rate profile measured beyond $10r_s$.

\section{Results}

In this paper, the mass inflow and outflow rates are defined as follows,

\begin{equation}
 \dot{M}_{\rm in}(r) = - 2\pi r^{2} \int_{0}^{\pi} \rho \min(v_{r},0)
   \sin \theta \rm d\theta,
   \label{inflowrate}
\end{equation}
\begin{equation}
 \dot{M}_{\rm out}(r) = 2\pi r^{2} \int_{0}^{\pi} \rho \max(v_{r},0)
    \sin \theta \rm d\theta,
    \label{outflowrate}
\end{equation}
and the net accretion rate is the sum of inflow rate and outflow rate.

\subsection{Mass inflow rate profile}

\begin{figure}
\begin{center}
\includegraphics[scale=0.45]{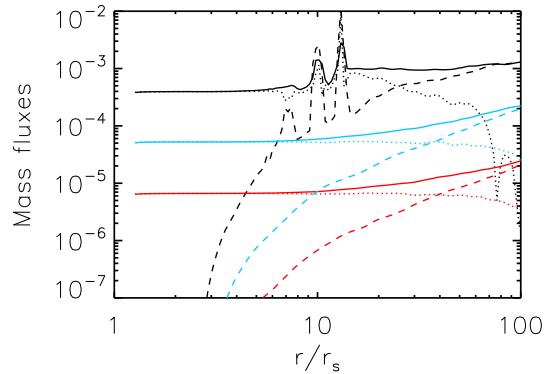}

\caption{ The radial profiles of the time-averaged (from $t=5$ to
7 orbits) and angle integrated mass inflow rate $\dot{M}_{\rm in}$
(solid lines), outflow rate $\dot{M}_{\rm out}$ (dashed line), and
the net rate $\dot{M}_{\rm acc}$ (dotted lines) in model A0 (black
lines), A1 (blue lines) and A2(red lines). The mass accretion rate is expressed in unit of Eddington accretion rate. \label{Fig:accretionrateA}}
\end{center}
\end{figure}

\begin{figure}
\begin{center}
\includegraphics[scale=0.45]{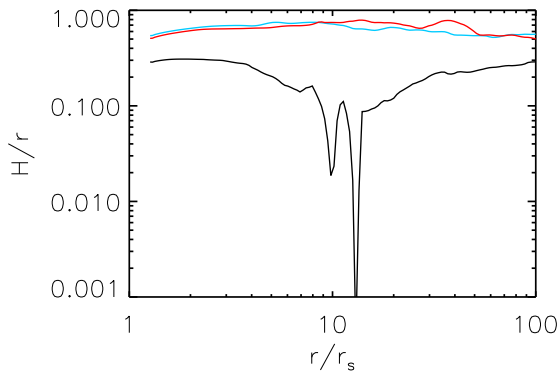}

\caption{ Time averaged (from $t=5$ to 7 orbits) scale height. The solution is averaged over an angle between $\theta=80^\circ$ to $\theta=100^\circ$. The black, blue and red lines correspond to models A0, A1 and A2, respectively. \label{Fig:heightA}}
\end{center}
\end{figure}

We take the models A0, A1 and A2 as our fiducial models. In these models, we set $\alpha=0.1$ and $\delta=0.1$.
If we turn on radiative cooling at the beginning of the simulation, the radiative cooling will be very strong at the torus center (because of the high density at the torus center). However, in the inner region, the radiative cooling is very weak because of the low density. This may induce non-physical results. In all the model, we first run the simulation without radiative cooling. We turn on the radiative cooling at 3 orbital times (measured at $200r_s$) when the accretion flows achieve a quasi-steady state.
Figure \ref{Fig:accretionrateA} shows the radial
profiles of the time-averaged mass inflow rate $\dot{M}_{\rm in}$,
outflow rate $\dot{M}_{\rm out}$ and net rate $\dot{M}_{\rm acc}$ in
models A0-A2. The mass accretion rate is expressed in unit of the Eddington accretion rate.

Figure \ref{Fig:heightA} shows the ratio of disc scaleheight to radius $H/r$, as a function of radius. The disc scaleheight is defined as $H\approx c_s/\Omega_k$, where $c_s$ is the sound speed calculated by using the ions pressure and $\Omega_k$ is the Keplerian angular velocity. The time averaged (from $t=5$ to 7 orbits) scale height is averaged over an angle between $\theta=80^\circ$ to $\theta=100^\circ$. The black, blue and red lines correspond to models A0, A1 and A2, respectively.

\begin{figure}
\begin{center}
\includegraphics[scale=0.45]{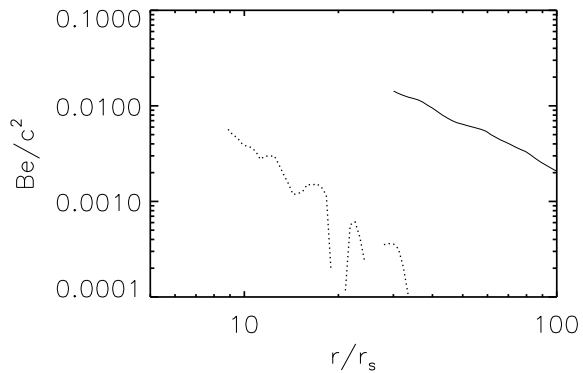}

\caption{ Radial distributions of time averaged (from $t=5$ to 7 orbits) and mass-flux weighted $Be$ in units of square of light speed. The dotted and solid lines are for models A0 and A2, respectively. \label{Fig:BeA}}
\end{center}
\end{figure}

\begin{figure}
\begin{center}
\includegraphics[scale=0.45]{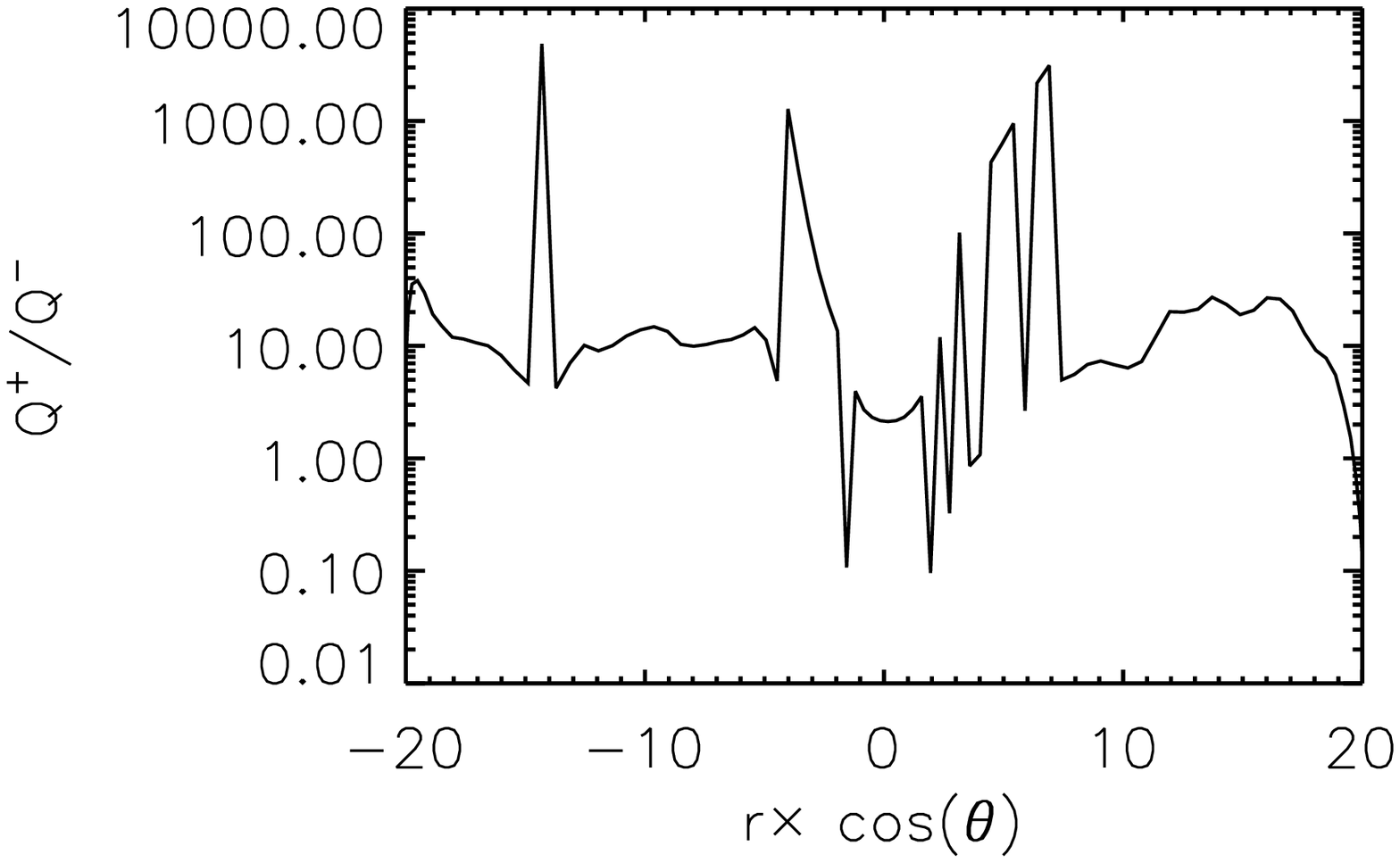}\\
\includegraphics[scale=0.45]{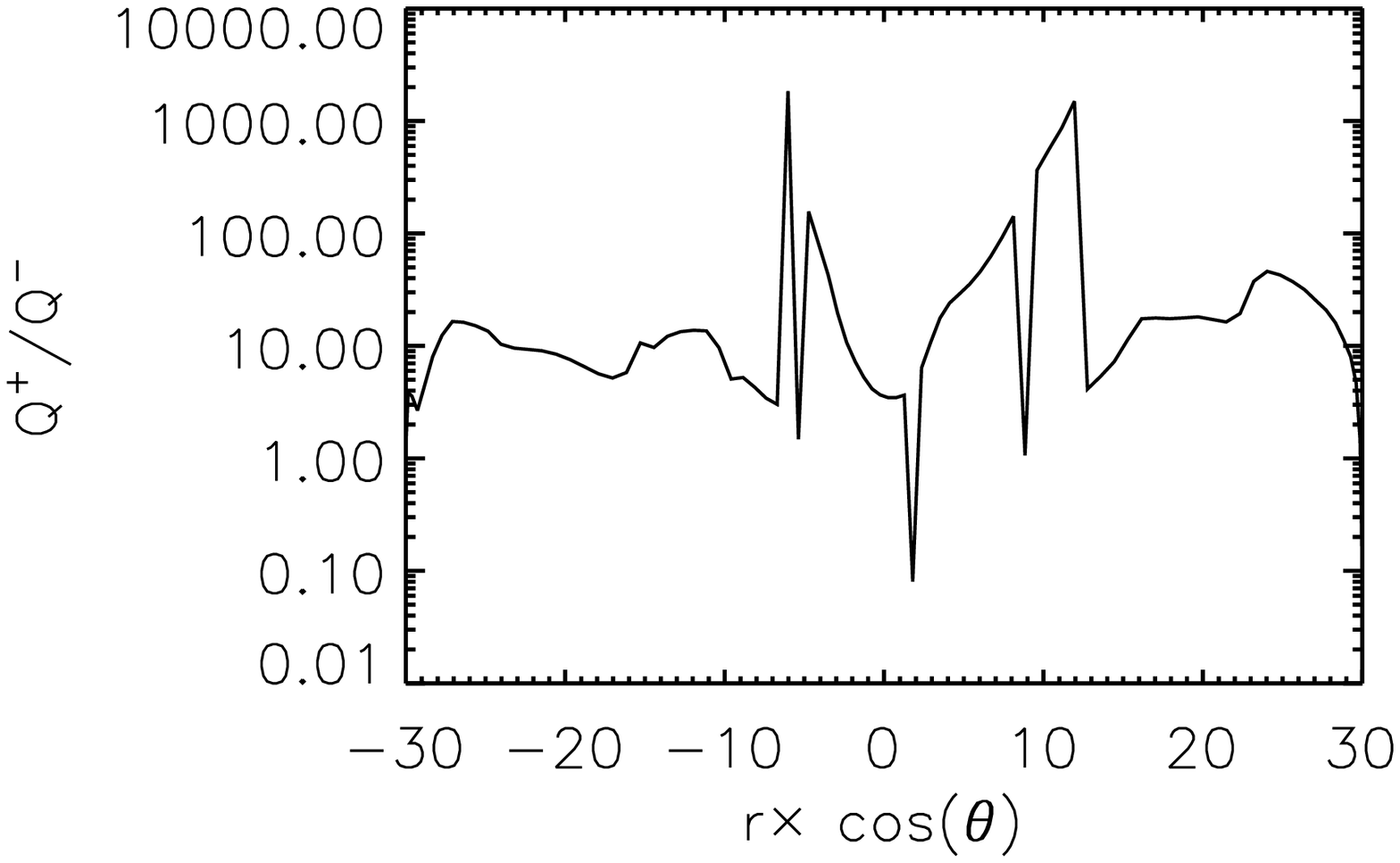}

\caption{ The $\theta$ dependence of the ratio of viscous heating rate to the radiative cooling rate at $t=7$ for model A0. Upper panel corresponds to $r=20r_s$. Bottom panel corresponds to $30r_s$. \label{Fig:chA}}
\end{center}
\end{figure}

In models A1 and A2, the net accretion rate is a constant of radius which indicates that our models have achieved quasi-steady state. In these two models, the mass accretion rate is very low, the radiative cooling is not important. The accretion flow is hot and thick. The aspect ratio ($H/r$) of the flow is larger than $0.5$ (see Figure \ref{Fig:heightA}).  In models A1 and A2, the mass inflow rate decreases inwards,
consistent with those found in previous works which study hot accretion flow without radiative cooling (see review in Yuan et al. 2012a). In the region $r> 10r_s$, the mass inflow rate in models A1 and A2 can be described by a power law function of radius $\dot {M_{\rm in}} \propto r^s$. The power law index of $s$ in models A1 and A2 is 0.54 and 0.57, respectively. This is also consistent with previous works (e.g. Yuan et al. 2012a). We have also done some test simulations with much smaller mass accretion rate than that in model A2, we find that the mass inflow rate profile in those models is identical with that in model A2.

Figure \ref{Fig:accretionrateA} shows that in models A1 and A2, there is a significant mass outflow rate. When we calculate the outflow rate, we include all the gas with positive velocity. Therefore, the mass outflow rate
includes both wind and gas which is doing turbulent
motions. Thus the crucial issue to quantitatively study the strength
of wind is to get rid of the contamination of turbulence. Yuan et al. (2015) develop ``trajectory method" to calculate the mass flux of wind. The details of this method can be found in Yuan et al. (2015).
Using the trajectory method, we find that the ratio of the mass flux of wind to total outflow rate calculated by Equation (\ref{outflowrate}) is $78\%$ in models A1 and A2. The ratio of wind mass flux to the inflow rate calculated by Equation (\ref{inflowrate}) is $55\%$. Therefore, the inward decrease of mass inflow rate in models A1 and A2 is due to mass loss via strong wind. This is consistent with that found in Yuan et al. (2015).

It is important to check whether the wind can escape to infinity. We calculate the Bernoulli parameter defined as:
\begin{equation}
Be=\frac{v^2}{2}+\frac{\gamma P}{(\gamma-1) \rho}-\frac{GM}{r-r_s}
\end{equation}
A positive $Be$ means that the wind has sufficient energy to overcome the gravitational energy and does work to its surroundings to escape to infinity. We take model A2 as an example for the low accretion rate case. Figure \ref{Fig:BeA} shows the radial distributions of time averaged (from $t=5$ to 7 orbits) and mass-flux weighted $Be$ in units of square of light speed. The dotted and solid lines are for models A0 and A2, respectively.  The Bernoulli parameter for wind in model A2 is positive. The wind can escape. We note that $Be$ decreases with increasing radius. This may be induced by radiative cooling. Due to the limited dynamical range of our simulation, we can not judge how $Be$ evolves when the wind moves to much larger radii. In future, it is necessary to investigate wind in a much larger dynamical range simulation.

\begin{figure}
\begin{center}
\includegraphics[scale=0.6]{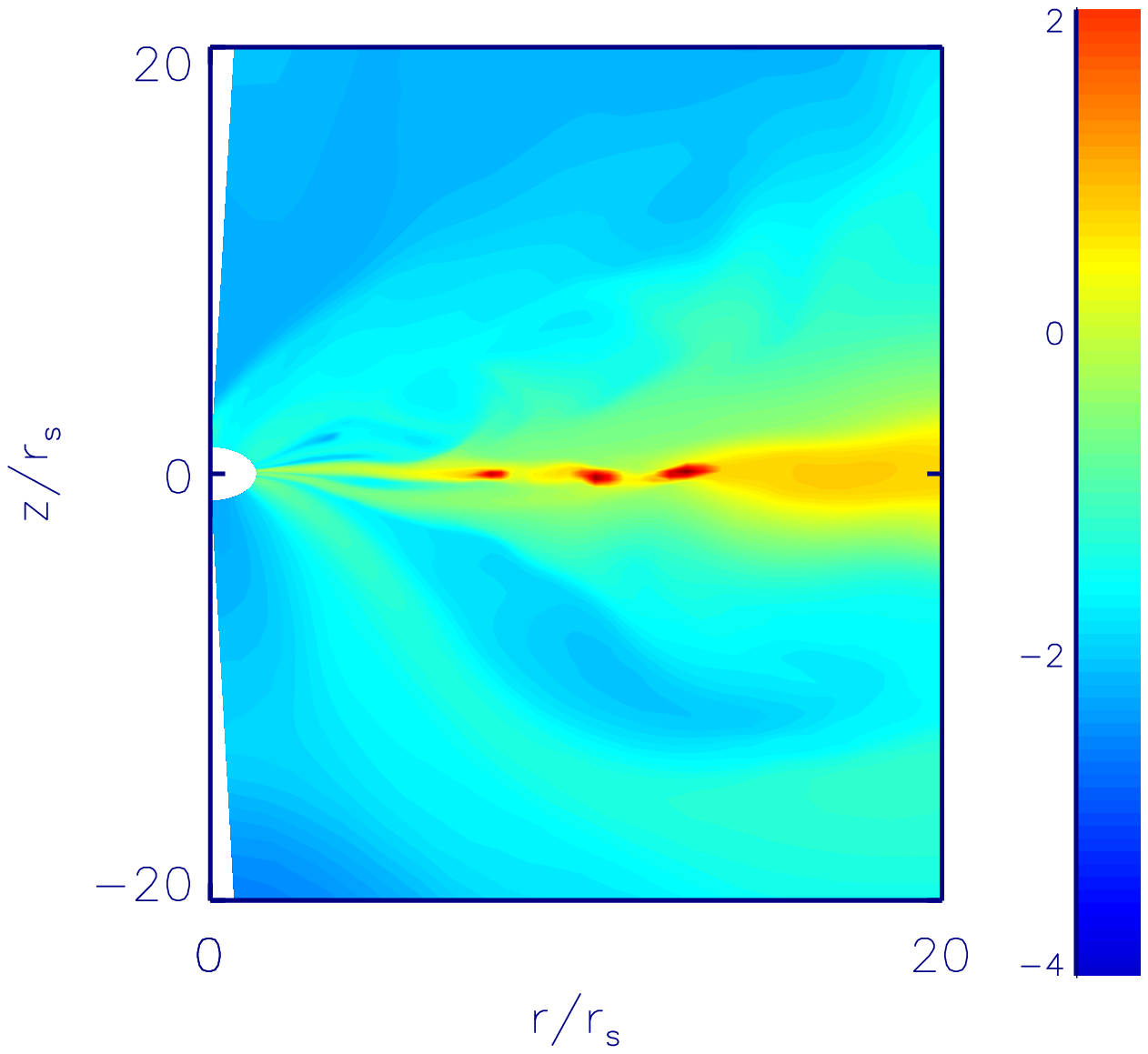} \\
\includegraphics[scale=0.6]{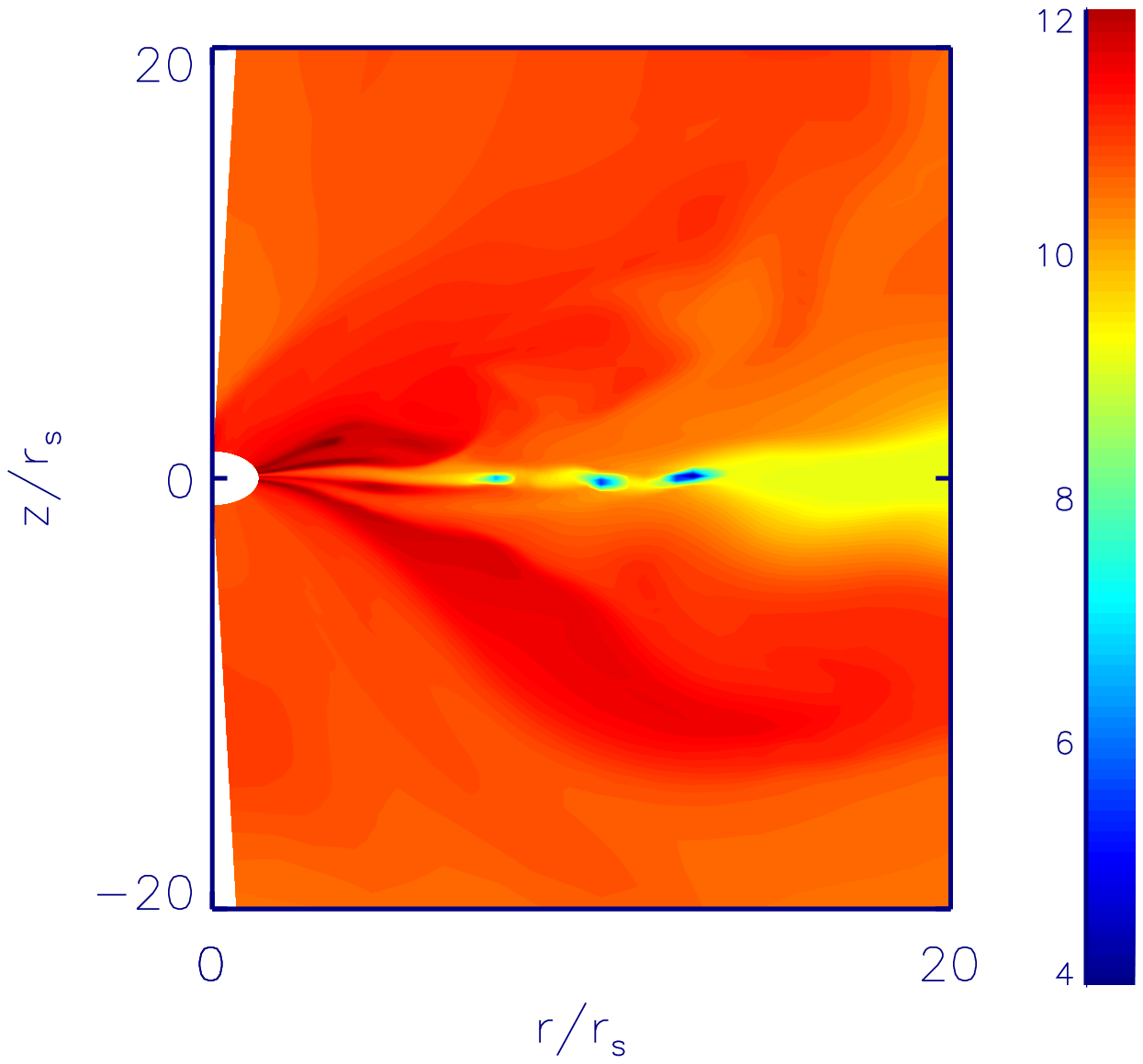}

\caption{Top panel: snapshot of the logarithm  $\log_{10}\rho/\rho_0$ ($\rho_0=7.8\times 10^{-16} \rm g/\rm {cm^{-3}}$) at $t=7$ orbits for model A0. Lower panel: snapshot of the logarithm temperature of ions at $t=7$ orbits for model A0. \label{Fig:snapdA0}}
\end{center}
\end{figure}

When the accretion rate increases, radiative cooling becomes more and more important. Figure \ref{Fig:chA} shows the $\theta$ dependence of the ratio between viscous heating and radiative cooling at $t=7$ for model A0. The upper and bottom panels correspond to $r=20r_s$ and $30r_s$, respectively. From this figure, we see that when the mass accretion rate is $4\times 10^{-4} \dot M_{Edd}$, at the region around the equatorial plane, the radiative cooling rate is bigger than the viscous heating rate. Radiative cooling effectively takes away energy and makes the accretion flow becoming geometrically thin (see Figure \ref{Fig:heightA} for the scale height). In the region $12 r_s < r < 100 r_s$, the mass inflow rate is almost a constant of radius. This is quite different from the case in which radiative cooling is not important. The net accretion rate (black dotted line in Figure \ref{Fig:accretionrateA}) is not constant of radius. This indicates that there is no steady state for accretion flow in model A0. Figure \ref{Fig:snapdA0} plots the snapshot (at $t=7$ orbits) of logarithm density and temperature for model A0. It is clear that in the region $10r_s < r < 20 r_s$, some cold clumps ($T \sim 10^4$ K) form. There are spikes in the mass accretion rate profile in region $10r_s < r < 20 r_s$. These spikes are induced by the transient accretion of cold clumps. Due to the presence of the spikes in the accretion rate profile, the net accretion rate in this model is not a constant of radius.

\begin{figure}
\begin{center}
\includegraphics[scale=0.45]{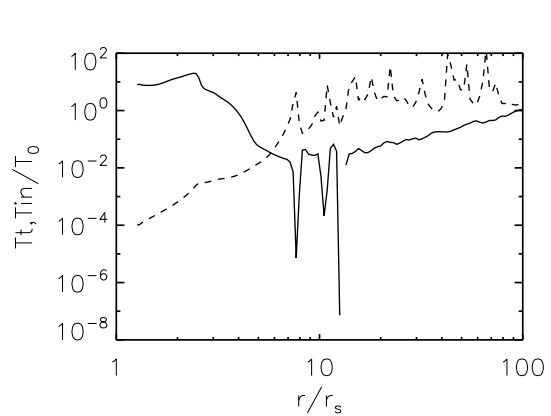}

\caption{ The radial dependence of the growth timescale of thermal instability (black line) and accretion timescale (dashed line) at $t=7$ for model A0. The timescales are calculated at the equatorial plane. The timescales are expressed in unit of the orbital time at $200r_s$. \label{Fig:timescaleA}}
\end{center}
\end{figure}

Yuan (2003) predicts that when the accretion rate of hot accretion flow is high, the thermal instability grow timescale can be shorter than the accretion timescale. In this case, strong radiative cooling may produce cold clumps and the flow may become multiphase. The growth timescale of thermal instability can be expressed as $T_t=2e_i/q^{ie}$ (Yuan 2003). The infall time scale of gas can be expressed as $T_{in}=r/v_r$.  Figure \ref{Fig:timescaleA} plots the infall and thermal instability timescales at the equatorial plane at $t=7$ for model A0. It is clear that, beyond $6r_s$, the growth timescale of thermal instability is more than one order of magnitude smaller than the infall timescale. This is the reason for the formation of cold clumps.

We note that when cold clumps form, the opacity of the gas in cold clumps may be high. In this case, it may be not appropriate to treat radiative cooling as optically thin, radiative transfer may need to be taken into account. Therefore, we can not use our simulation to give the quantitative properties of cold clumps (e.g., covering factor, mass ratio between cold and hot gas). In future, we plan to perform simulations with radiative transfer to investigate the properties of two-phase accretion flow.

\begin{figure}
\begin{center}
\includegraphics[scale=0.6]{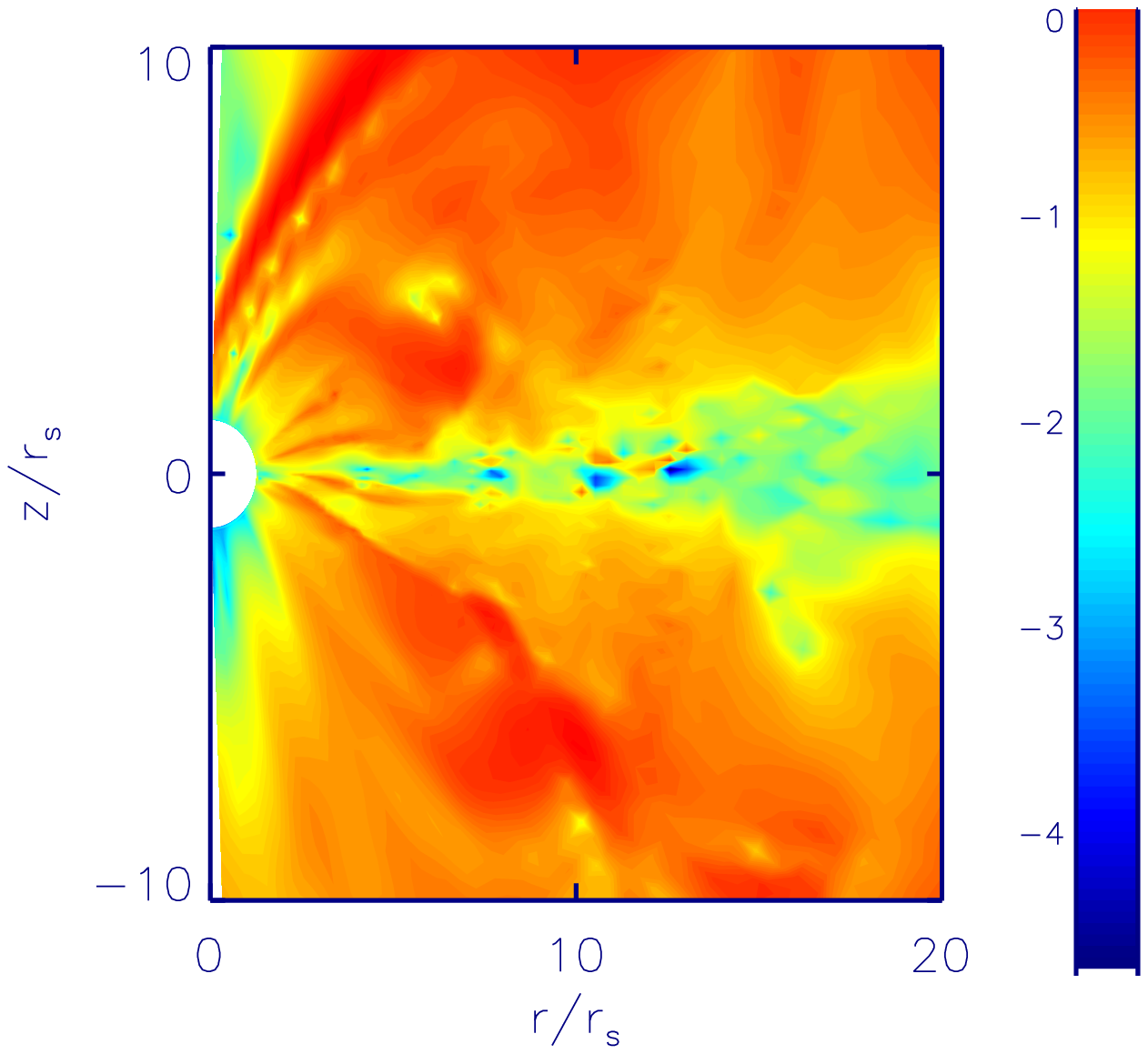} \\
\includegraphics[scale=0.6]{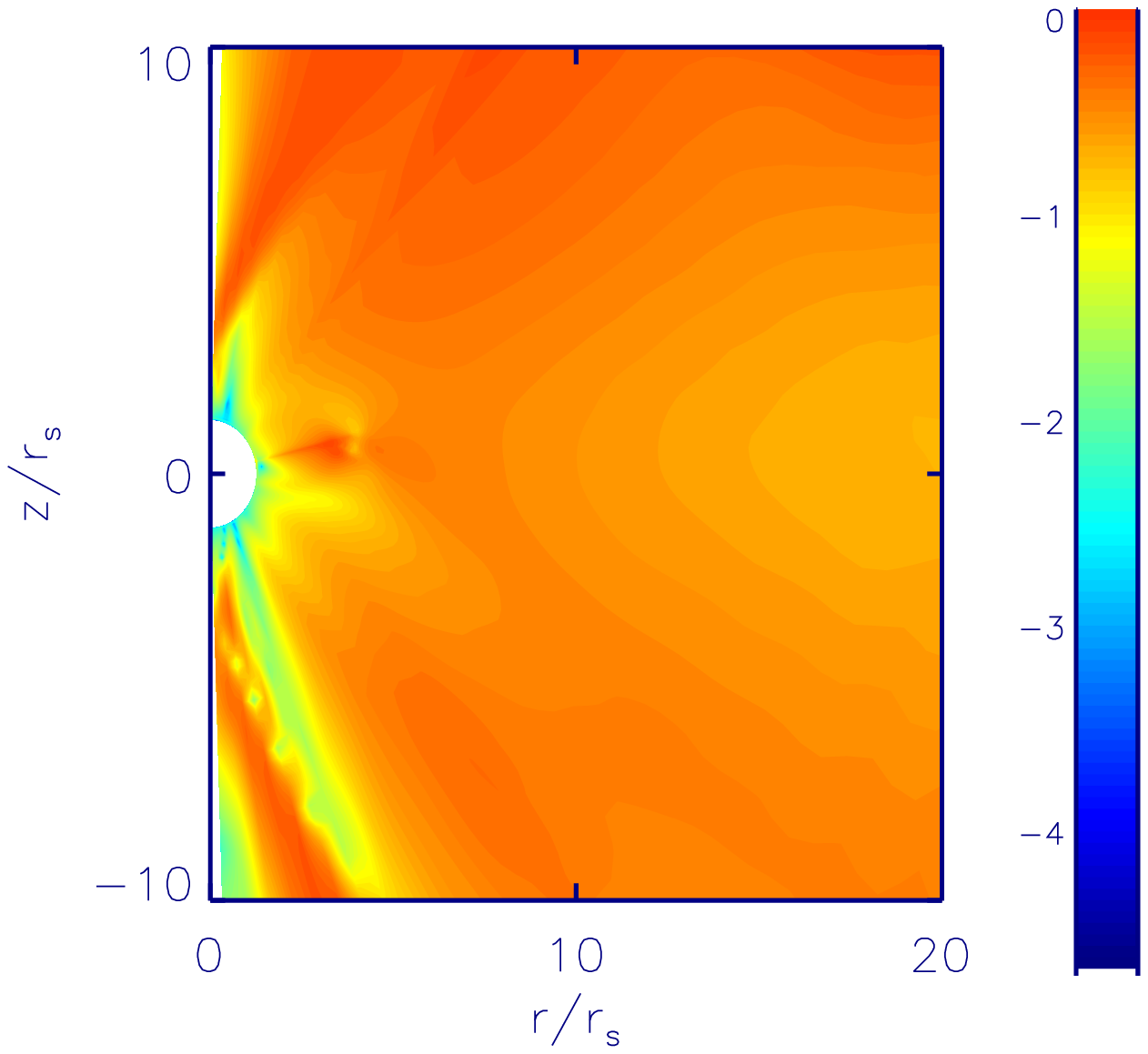}

\caption{Top panel: snapshot of the $\log_{10}\frac{\nabla p}{\rho GM/(r-r_s)^2}$ at $t=7$ orbits for model A0. Lower panel: snapshot of the $\log_{10}\frac{\nabla p}{\rho GM/(r-r_s)^2}$ at $t=7$ orbits for model A2. \label{Fig:forceA0}}
\end{center}
\end{figure}

Using the trajectory method, we find that the ratio of the mass flux of wind to total outflow rate calculated by Equation (\ref{outflowrate}) is $5\%$ in model A0. Wind is very weak in model A0. In the HD simulation, wind can be accelerated by the combination of gas pressure gradient and centrifugal forces (Yuan et al. 2012b). Radiative cooling takes away internal energy of gas and makes the pressure becoming small. This is the reason for the weakness of wind when radiation becomes important. We calculate the ratio of pressure gradient force to gravitational force for models A0 and A2 and show the results in Figure \ref{Fig:forceA0}. It is clear that in model A0, especially in the region close to the equatorial plane, the gas pressure gradient force decreases significantly when radiation becomes important. We also note that density close to the equatorial plane is much larger than that at small $\theta$ angles. Therefore, the decrease of pressure gradient force close to the equatorial plane can significantly decrease the mass flux of wind.

We also calculate the Bernoulli parameter for wind in model A0 (dotted line in Figure \ref{Fig:BeA}). In the region $r< 30r_s$, $Be$ is positive. However, when the wind moves to the region beyond $30r_s$, $Be$ becomes negative due to radiative cooling. Therefore, the wind in model A0 can not escape to infinity.

Yuan \& Bu (2010) studied the wind strength in presence of radiative cooling. They find that when $\dot{M}_{\rm BH}\sim 7\times 10^{-4}\dot M_{Edd}$, wind is still very strong. We note that in that work only bremsstrahlung radiation is taken into account. In the region beyond $30r_s$ the radiative cooling rate is higher than viscous heating rate. However, in the region $r< 30 r_s$, viscous heating rate is larger than radiative cooling rate. In this paper, in addition to bremsstrahlung radiation, we have synchrotron emission and the Comptonization. When $\dot{M}_{\rm BH}\sim 4\times 10^{-4}\dot M_{Edd}$, radiative cooling rate can dominate viscous heating rate even in the inner region $r< 30 r_s$. This makes the wind mass flux decreasing significantly.

\begin{figure}
\begin{center}
\includegraphics[scale=0.45]{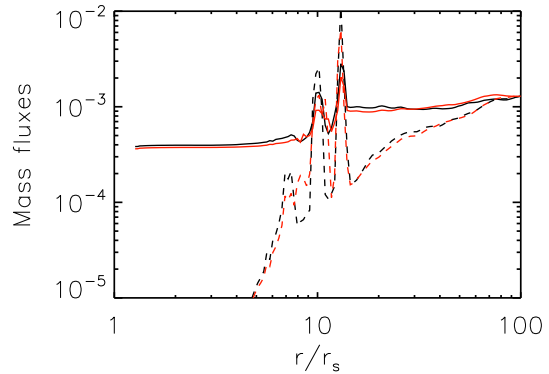}

\caption{ The radial profiles of the time-averaged (from $t=5$ to
7 orbits) and angle integrated mass inflow rate $\dot{M}_{\rm in}$
(solid lines) and outflow rate $\dot{M}_{\rm out}$ (dashed line) in model A0 (black
lines) and A0h (red lines). The mass accretion rate is expressed in unit of Eddington accretion rate. \label{Fig:accretionrateA0h}}
\end{center}
\end{figure}

\subsection{Dependence on different parameters}
The initial torus is embedded in a low density background gas. The density of background gas should be low enough to guarantee that at the torus surface there are no artificial shocks or oscillations induced by artificial pressure gradient (Rezzolla et al. 2003a, 2003b; Zanotti et al. 2003, 2005; Blaes et al. 2006; Mishra et al. 2017). If there are shocks and oscillations at the initial torus surface, the results discussed in this paper may be affected. We carry out model A0lm to check this point. The only difference between models A0lm and A0 is the density of the background gas. The density of background gas in model A0lm is 10 times smaller than that in model A0. We find that the results in model A0lm are almost same with those in model A0. Therefore, our results do not depend on the background gas density.

In order to test whether the results depend on the resolution of simulations, we carry out model A0h. The only difference of models A0 and A0h is the resolution. The resolution of simulation in model A0h is two times of that in model A0. Figure \ref{Fig:accretionrateA0h} shows the radial profiles of the time-averaged mass fluxes in
models A0 and A0h. In this figure, the black and red lines correspond to models A0 and A0h, respectively. The results of models A0 and A0h are very similar. We quantitatively calculate the wind mass flux in model A0h. We find that the wind mass flux is $4\%$ of the mass outflow rate calculated by Equation (\ref{outflowrate}). In model A0, this ratio is $5\%$. The resolution used in this paper is enough to study the strength of wind.

In the simulations introduced above, we assume that $\delta=0.1$. Ten percent of the viscous heating energy directly goes into electrons. At present, investigations show that $\delta$ lies in the range $0.1-0.5$ (see Yuan \& Narayan 2014 for reviews). If $\delta$ is larger, more viscous energy directly goes into electrons resulting in stronger radiative cooling. It may affect the results discussed above. In order to study the effects of $\delta$, we carry out models B0, B1 and B2 with $\delta=0.5$.

We find that when the mass accretion rate (measured at the inner boundary) is small ($\dot M < 10^{-4} \dot M_{\rm Edd}$), radiative cooling is not important and the flows are hot accretion flows (models B1 and B2). In models B1 and B2, in the region $r> 10r_s$, the mass inflow rate can be described by a power law function of radius $\dot {M}_{\rm in} \propto r^s$. The power law index of $s$ in models B1 and B2 is 0.56 and 0.53, respectively. Wind is very strong in these two models. The ratio of the mass flux of wind to total outflow rate calculated by Equation (\ref{outflowrate}) is $70\%$. The ratio of wind mass flux to the inflow rate calculated by Equation (\ref{inflowrate}) is $50\%$.

When the mass accretion rate is high (model B0) $ \dot {M}=3.3\times 10^{-4} \dot M_{\rm Edd}$, we find that at the region around the equatorial plane, some cold clumps ($T \sim 10^4$ K) form around the equatorial plane. In this model, wind is very weak. We find that the ratio of the mass flux of wind to total outflow rate calculated by Equation (\ref{outflowrate}) is $4\%$. The accretion rate at which cold clumps form in model B0 is smaller than that in model A0. This is because $\delta$ in model B0 is bigger, this induces stronger radiative cooling and makes the accretion rate at which cold clumps form in model B0 lower.

In order to test the dependence of results on viscous coefficient $\alpha$, we carry out models C0, C1 and C2. In these models, we set $\alpha=0.01$, which is one order of magnitude smaller than that in models A and B series. In model C series, we set $\delta=0.1$. Because $\alpha$ is small in model C series, the simulations need much longer time to achieve quasi-steady state. We find that the quasi-steady state is achieved after 30 orbital times (measured at $200r_s$). We turn on radiative cooling at 30 orbital times in models C0, C1 and C2.

When the mass accretion rate (measured at the inner boundary) is small ($\dot M < 10^{-5} \dot M_{Edd}$), the radiative cooling is not important and the flows are hot. The wind is very strong. The ratio of the mass flux of wind to total outflow rate calculated by Equation (\ref{outflowrate}) is $73\%$.  The ratio of wind mass flux to the inflow rate calculated by Equation (\ref{inflowrate}) is $51\%$. When the mass accretion rate is $2.2\times 10^{-5} \dot M_{Edd}$, a cold thin disk forms around the equatorial plane. In model C0, where a cold disk forms at the equatorial plane, the accretion rate is smaller than that in model A0 by a factor of $20$. This is because, in model C0 $\alpha$ is smaller than that in model A0 by one order of magnitude, therefore, the viscous heating is much smaller in model C0. We find that the ratio of the mass flux of wind to total outflow rate calculated by Equation (\ref{outflowrate}) is $6\%$ in model C0.

\section{Summary and discussion}
In this paper, we perform two-dimensional hydrodynamical simulations to study how wind strength changes with mass accretion rate onto the black hole. We solve the energy equation of ions and electrons separately. We take into account bremsstrahlung, synchrotron radiation and the Comptonization.  When the accretion rate is low, radiative cooling is not important, the accretion flow is hot accretion flow. For the hot accretion flow, the mass inflow rate can be described by a power law function of radius $\dot {M}_{\rm in} \propto r^s$ with $0.5<s<0.6$. The inward decrease of mass inflow rate is due to mass loss via strong wind.

With the increase of accretion rate, radiative cooling becomes important. We find that when radiative cooling rate is bigger than the viscous heating rate, a cold thin disk or cold clumps around the equatorial plane will form. In this case, the mass inflow rate can not be described by a power law function of radius. Also, wind is very weak. The decrease of wind mass flux is due to the decrease of gas pressure gradient force, which can accelerate wind. We need to note that when cold clumps or cold disk form, the optical depth may be high and the radiative transfer may be needed. When radiative transfer is included, the radiation pressure may help to drive wind. We will study this point in future.

We also need to note that the accretion rate at which cold thin disk or cold clumps can form is not accurate. This is because in this paper we have two many simplifications about the real physics. For example, we do not take into account radiative transfer and magnetic field effects. We can just conclude that when the accretion rate is small and radiative cooling rate is much smaller than viscous heating rate, wind is very strong. The exact value of the mass accretion rate at which cold thin disk or clumps form need to be studied in details in future. Below we will highlight several notable shortages of this work and discuss their possible consequences.

The Comptonization is local Compton scattering. However, in hot accretion flow with radial optical depth less than unity, the photons generated at small radii can go for a large distance and heat or cool gas at large radii. Such 'global' Comptonization is not taken into account. Global Compton scattering may influence the properties of accretion flow significantly (e.g. Ostriker et al. 1976; Cowie et al. 1978; Park \& Ostriker 2007; Yuan et al. 2009; Xie et al. 2010). At large radii, such Compton scattering can heat the gas; the gas at large radii may be heated above the virial temperature and thus forms outflows. This may change wind strength.

In accretion flows, ordered, large-scale open magnetic field may exist (e.g. Blandford \& Payne 1982; Lovelace et al. 1994; Cao 2011; Penna et al. 2013; Bai \& Stone 2013; Li \& Begelman 2014). Large scale magnetic field can affect the dynamics of the flow by two ways. First, magnetic field can transfer angular momentum by magnetic braking effect (Stone \& Norman 1994). Second, magnetocentrifugal wind can be launched by large scale magnetic field (Blandford \& payne 1982; Cao 2011; Li \& Begelman 2014). If magnetocentrifugal wind exists, the wind strength may be changed.

In the cosmological simulations studying galaxy formation and evolution (e.g., Springel et al. 2005; Gan et al. 2014), the Bondi radius can at most be marginally resolved. The authors use sub-grid models to study the black hole growth and AGNs feedback. The Bondi accretion model is usually used to estimate the accretion rate of the black hole. According to our work, when the accretion flow is hot accretion flow, wind is very strong. Therefore, the Bondi formula over-estimates the accretion rate to the black hole. In the cosmological simulations, authors can use the Bondi formula to calculate the accretion rate at the Bondi radius. Then, they can use a power law function to calculate the decrease of the mass accretion rate from Bondi radius to the black hole. The power law index is around 0.5.

\section*{Acknowledgments}
This work is supported in part by the National Program on Key Research and Development Project
of China (Grant No. 2016YFA0400704),  the Natural Science Foundation of China (grants
11573051, 11633006, 11773053 and 11661161012), the Natural Science
Foundation of Shanghai (grant 16ZR1442200), and the Key
Research Program of Frontier Sciences of CAS (No. QYZDJSSW-
SYS008).  This work made use of the High Performance Computing Resource in the Core
Facility for Advanced Research Computing at Shanghai Astronomical
Observatory.

\end{document}